\documentstyle[sprocl]{article}

\def\Journal#1#2#3#4{{#1} {\bf #2}, #3 (#4)}

\def\PRL{\em Phys. Rev. Lett.}
\def\PRD{{\em Phys. Rev.} D}
\def\ZPC{{\em Z. Phys.} C}

\def\be{\begin{equation}}
\def\ee{\end{equation}}
\def\bea{\begin{eqnarray}}
\def\eea{\end{eqnarray}}

\begin{document}

\title{Results from Searching for  Muonium to Antimuonium Conversion }

\author{K. JUNGMANN, A. GROSSMANN, J. MERKEL, V.MEYER,
G.~ZU~PUTLITZ, I.~REINHARD, K. TR\"AGER, 
P.V. SCHMIDT, L. WILLMANN}

\address{Physics Institute, University of Heidelberg, D-69120 Heidelberg }

\author{R. ENGFER, H.P. WIRTZ}

\address{Physics Institute, University of Z\"urich, CH-8057 }

\author{R. ABELA, W. BERTL, D. RENKER, H.K. WALTER}

\address{Paul Scherrer Institute; CH-5232 Villigen}

\author{V. KARPUCHIN, I. KISEL, A. KORENCHENKO, 
S. KORENCHENKO, N.~KRAVCHUK, N. KUCHINSKY, A. MOISEENKO}

\address{JINR Dubna, RU-141980 Dubna}

\author{J. BAGATURIA, D. MZAVIA, T. SAKHELASHVILI}

\address{Tbilisi State University, GUS-380086 Tbilisi}

\author{V.W. HUGHES}

\address{Yale University, New Haven, CT 06520}


\maketitle
\abstracts{ A new result from searching for muonium to antimuonium conversion is reported
 which sets an upper limit on the coupling constant in an assumed (V-A)$\times$(V-A)
 type interaction of G$_{M\overline{M}} \leq 3\cdot 10^{-3}{\rm G_F}$~ (90\%C.L.).
 A particular Z$_8$ and a minimal 331 GUT model can be ruled out. Further new and stringent
 limits can be set for masses of bileptonic gauge bosons and $\lambda$ parameters in
 R-parity breaking supersymmetric models.}


Due to the close confinement in the bound state  
the hydrogen-like muonium atom ($M=\mu^+e^-$) is a well suited 
system to test the fundamental interactions between
two leptons from different generations. 
Of particular interest are searches for lepton 
number violating processes which occur naturally in many 
speculative extensions
to the standard model.
A spontaneous conversion of muonium to
antimuonium ($\overline{M}=\mu^-e^+$) would 
violate additive lepton number conservation by two units and is of interest
particularly to  e.g. left-right
symmetric \cite{Her_92,Hal_82}, supersymmetric \cite{Moh_92,Hal_93} and GUT
\cite{Fuj_94,Hou_96,Fram_97} models. 
\\

At the Paul Scherrer Institute (PSI) in Villigen, Switzerland, 
an experiment has been successfully completed very recently 
which searched sensitively for $M-\overline{M}$ conversion. 
The coincident detection of both of the
constituents ($\mu^-$ and $e^+$) of the antiatom in its decay 
was required as a signature for the conversion process \cite{Abela_96}.

A $\mu^-$--decay
($\mu^-\rightarrow e^-+\overline{\nu}_e+\nu_{\mu}$)
releases an energetic electron with an energy spectrum ranging up to 53 MeV
and can be identified in a magnetic
spectrometer consisting of five cylindrical
proportional wire chambers in coaxial geometry operated at 0.1~T field.
The positron in the atomic shell
of the antiatom is left behind after the decay with an average
kinetic energy of $R_{\mu}$=13.5~eV. This particle
is accelerated to 7~keV and transported
in a magnetic guiding field onto a position sensitive microchannel plate
detector (MCP). The 511~keV radiation from the annihilation
of positrons in the MCP
can be detected in a segmented pure CsI crystal detector.

The muonium atoms are formed by stopping
a beam of positive muons
in a $ {\rm SiO_2}$ powder.
A fraction of about 7\% of the incoming muons
is converted into muonium atoms which diffuse to the surface and leave the
powder with thermal energies into the surrounding vacuum.
It is essential for the experiment to have the atoms in vacuum,
since the $M$--$\overline{M}$-oscillation is strongly suppressed
in gases or condensed matter due to the removal of symmetry
between the atom and the antiatom.
The muonium yield has been monitored regularly
by reversing the electric acceleration field and all magnetic fields
approximately every five hours
and by observing energetic positrons and electrons respectively.
These measurements also serve for calibrating detector subsystems.
The targets are replaced about every third day to account for an
observed time dependent decrease of the muonium production efficiency,
due to target surface structure and contaminations.
Attention has been paid to keep the apparatus as much as possible symmetric
for detecting muonium and antimuonium decays.

In the analysis of the final data taking period, which included 1290h 
of apparatus lifetime, 
one event
was found within 3 standard deviations of all relevant parameters. This is in 
good agreement with an expected background of 1.7(2) events from accidental 
coincidences. The physical background due to the allowed process
$\mu^-\rightarrow e^-+\overline{\nu}_e+\nu_{\mu} + e^+ + e^-$ 
which could fake an event is fully suppressed
by a proper choice of the magnetic positron transport system parameters 
which includes a 90$^\circ$ bend. 
The combination of all measurement periods encompassing some 1600 h measurement time
yields an upper limit on the conversion probability 
of $P_{M\overline{M}}$ $\leq$ $8.0 \cdot 10^{-11} / S_b$ (90\% C.L.),
where $S_b$ reflects the magnetic field dependence of the process and
is for 0.1~T in this experiment 0.35 for (V$\pm$A)$\times$(V$\pm$A) and 0.75 
for (V$\mp$A)$\times$(V$\pm$A) interaction types \cite{Wong_95}.
This constitutes an improvement by a  factor of 2500 over 
the value established  in preceding experiments \cite{Matt_91}. 
This corresponds for an assumed effective four fermion in (V-A)x(V-A)  
interaction to a limit on the coupling constant of 
G$_{M\overline{M}}$ $\leq$ $3 \cdot 10^{-3}$ G$_F$, 
with G$_F$ being the weak interaction Fermi coupling constant.
\\


This result rules out a $Z_8$ model with
radiative mass generation by heavy lepton seed \cite{Hou_96}.
It  allows further to set
a lower limit on the mass of bileptonic gauge bosons in GUT theories
of 2.6~TeV/c$^2$ * g$_{3l}$, where  g$_{3l}$ depends on the particular
symmetry and is of order unity; this is significantly 
above values extracted from high energy Bhabha
scattering experiments. In a minimal 331 model the mass limit on the bilepton
an 800 GeV/c$^2$ upper mass bound has been found. The new  limit from this
${M\overline{M}}$ experiment
corresponds to a 850 GeV/c$^2$ lower bound and excludes the minimal
version of this theory. However, the model would still be viable, if one 
add an octet of Higgs \cite{Fram_98}.

In the framework of R-parity violating supersymmetric models 
\cite{Moh_92,Hal_93}the bound
on the coupling parameters could be lowered by a factor of 15 
to $\mid~\lambda_{132}\lambda_{231}~\mid \leq 3* 10^{-4}$ with assumed
superpartner masses of order 100 GeV/c$^2$ \cite{Moh_98}.
The achieved level of sensitivity allows to narrow slightly the 
interval of allowed 
heavy muon neutrino masses
in minimal left-right symmetric
models \cite{Her_92} (which have predicted a lower bound on
${\rm G_{M\overline{M}}}$ provided muon neutrinos have a mass above
35 keV) to $\approx$ 40~keV/c$^2$ up to the present
experimental bound at 170~keV/c$^2$.
\\

The work is supported in part by the BMBF of Germany, the Russian RFFR
and the Schweizer Nationalfond and a NATO research grant.

\end{document}